\documentstyle[12pt,psfig]{article}
\begin{document}
\begin{titlepage}
\begin{center}
{\large \bf {Statistical model for pionic partons}}
\end{center}
\centerline{\bf N. G. Kelkar$^{a,}$\footnote{email 
address: ngkelkar@apsara.barc.ernet.in}
and M. Nowakowski$^{b,}$\footnote{email 
address: marek@marik.uniandes.edu.co}
}
\vskip0.25cm
\centerline{$^a$ \it Nuclear Physics Division,
Bhabha Atomic Research Centre,} 
\centerline{\it Mumbai 400 085, India.}
\centerline{$^b$ \it Departamento de Fisica, Universidad de Los Andes}
\centerline{\it A.A.4976, Bogota D.C., Colombia}
\begin{abstract}
We present a model for the structure of the pion. 
Based on ideas of a recently developed statistical model of the
nucleon, we assume the pion to be a gas of partons. 
The finite-size corrections (FSC) are incorporated through 
two parameters. Using the same two FSC 
parameters for the proton and pion we reproduce 
quantitatively the data on $\pi^- N \rightarrow \mu^+ \mu^- X$ Drell-Yan
production and valence quark distribution of the pion. 
\end{abstract}
\end{titlepage}
\newpage
\section{Introduction}
Recently, a statistical model for the structure of the nucleon was
proposed \cite{bhal}. The nucleon was described as a non-interacting 
gas of valence quarks, sea quarks and gluons in equilibrium. This 
picture was then improved by adding finite-size corrections (FSC) to
the expression for the parton density function (PDF). 
These finite size corrections take into account the fact that the gas
is enclosed in a finite volume.
The model was  
successful in reproducing a large body of polarized and 
unpolarized nucleon structure function data. Since the ideas from 
statistical mechanics worked well inside one hadron, namely the
nucleon, it is quite motivating to test these ideas in other hadrons as well. 
However, compared to the vast amount of precise data available for
the nucleon, much less is available for others. Among hadrons other
than the nucleon, the pion is the most widely explored. There 
exist data on $\pi^- N \rightarrow \mu^+ \mu^- X$ Drell-Yan production
\cite{na10,e615} and prompt photon production in $\pi^{\pm} p \rightarrow
\gamma X$ \cite{prompt} which have been used by some groups 
\cite{smrs,grs} to obtain parameterizations for PDFs of the pion.
The Drell-Yan data constrains the shape of the pion valence densities
and the prompt photon data constrains the pionic gluon distribution
in the large-x region. These data are, however, not sufficient to fix the
gluon and sea in the pion uniquely. The number of parameters appearing
in the expressions for the PDFs obtained from such global fits are very
large and with no physical meaning. Hence it is worth investigating
the scope of a statistical model which has fewer parameters 
based on a physical interpretation.

In this Letter, we present a statistical model for the parton 
distributions in the pion. 
The expressions
for the pion PDFs are written assuming the pion to be a gas of
massless partons. These are then improved to include FSC as in 
the case of the
nucleon. In principle the two parameters associated with the FSC
can be determined theoretically \cite{paras}. However, these
parameters are sensitive to the equation of motion employed,
the boundary conditions on the wave function, shape of the enclosure
and details such as whether the particles are strictly massless. 
In ref.\cite{bhal}, in
the absence of a complete understanding of the nucleon as a QCD
bound state, a practical approach was taken and 
the two parameters for the nucleon were determined
by fitting unpolarized structure function $F_2(x,Q^2)$ data at
one value of $Q^2$. In the case of the pion, due to scarcity of 
precise experimental data giving direct information on the pion PDFs,
we prefer to use as a first guess, the same parameters for the pion as
in the case of the proton \cite{bhal}. Interestingly, with 
these parameters we
get good quantitative agreement with the available data on pion
valence quark distribution and $\pi^- N \rightarrow \mu^+ \mu^- X$ 
Drell-Yan production. The agreement with data is nearly as good as
that obtained by the existing parameterizations \cite{smrs,grs}. 

\section{The model}
We visualise the pion to be made up of a gas of massless partons in
equlibrium at temperature $T$ in a spherical volume $V$ with radius $R$. The
parton number density $dn^i/dx$ in the infinite momentum frame (IMF) and
the density $dn/dE$ in the pion rest frame are related to each other
by, 
\begin{equation}
{dn^i \over dx}
= {M^2 x \over 2} \int ^{M/2}_{xM/2} {dE \over E^2}~~ {dn \over dE},
\end{equation}
where the superscript $i$ refers to the IMF, $M$ is the pion mass
and $E$ is the parton energy in the pion rest frame \cite{bhal,rsb}. 
Using the standard procedure in statistical mechanics to introduce the 
effects of the finite size of an enclosure, in the expression for
the density of states \cite{hw} we write $dn/dE$ for the pion as,
\begin{equation}
dn/dE = g ~f(E)~ (VE^2/2\pi^2 + aR^2E + bR),
\end{equation}
where $g$ is the spin-color degeneracy factor, $f(E)$ is the usual
Fermi or Bose distribution function $f(E) = \{\exp[(E-\mu)/T]\pm
1\}^{-1}$, $V$ is the pion volume and $R$ is the radius of a sphere
with volume $V$. We take $R\, = \,\sqrt{5/3} \rho \, =\, 1.07 fm$, 
where $\rho$ is the root-mean-square radius of the pion \cite{pirad}.
The three terms in (2) are the volume, surface and
curvature terms, respectively. The coefficients $a$ and $b$ are, 
in principle, the 
free parameters of the model. For reasons mentioned in the introduction,
though $a$ and $b$ can be determined theoretically, we prefer to choose
these values to be the same as those determined phenomenologically
for the proton in ref.\cite{bhal}. Theoretically this makes sense, 
as the values
$a = -0.376$ and $b = 0.504$ obtained from the fitting procedure in
ref.\cite{bhal} are close to 
the values ($a=-1/2$ and $b=3/2\pi$) determined theoretically by
Morse and Ingard \cite{paras}. We will come back to this point at the 
end of the paper.

To determine the values of the temperature and chemical potentials
appearing in the distribution function $f(E)$, we consider the
fact that any model of the PDFs for the pion has to obey the number and
momentum constraints, the former being motivated by the constituent 
quark picture for mesons
at a low scale (this scale, being our input scale, 
should not be to too small as compared with $\Lambda_{QCD}$ to allow for
the evolution of the partons densities, hence we choose it to be 
$Q_0^2=m_N^2$ where $m_N$ is the mass of the nucleon). 
If $n_{\alpha(\bar \alpha)}$ denotes the number
of quarks (antiquarks) of flavour $\alpha$, then in the case of a 
$\pi^-$ for example, the constraints can be given as,
\begin{eqnarray}
n_{\bar u} \, - \, n_u \, = \, 1 \\
n_d \, - \, n_{\bar d} \, = \, 1 \\
n_s \, - \, n_{\bar s} \, = \, 0 \\
\sum_{q,\,\bar{q}, \, g}\, (momentum fractions)\, = \, 1. 
\end{eqnarray}
The numbers $n_{\alpha(\bar \alpha)}$ in eqs.(3-5) are obtained from
eqs.(1) and (2) by integrating the appropriate $dn^i/dx$ over $x$. 
The momentum fractions are obtained by integrating 
$x dn^i/dx$ over $x$.
The temperature $T$ and chemical potentials ($\mu$) appearing in eqs.(3-6)
are then not free parameters. For a given set of parameters $a$ and $b$, they
are determined uniquely by solving the four coupled non-linear
eqs.(3-6).

Using the values of $a$ and $b$ mentioned above, we obtain $T = 22$ MeV, 
$\mu_d = -\mu_u = 162$ MeV and $\mu_s = 0$ in the case 
of a negatively charged pion. 
The left and right hand sides of eqs.(3-6) with these values
of T and $\mu$, agree with each other up to an accuracy of one part
in $10^6$. The chemical potentials for antiquarks are determined by 
using the relation $\mu_{\bar q} = - \mu_q$. The model, 
described above, fixes the parton distributions
in a pion uniquely at an input scale $Q_0$. It remains to put it to test 
and compare the predictions with available 
experimental data. In order to compare with data, the parton distributions
are evolved to different $Q^2$ values using the standard DGLAP 
evolution equations at next-to-leading
order, starting from the input scale $Q_0^2 = m_N^2$.

\section{Results and discussion}
To start with, we compare the $\pi^-$ valence quark distribution 
calculated within the statistical model with the available data
at $Q^2 = 30$ GeV$^2$. In Fig. 1a we show the valence structure function
extracted from $\pi^- W \rightarrow \mu^+ \mu^- X$ Drell-Yan data by
the E615 collaboration.  The solid curve corresponding to our
calculation of $x v_{\pi} = x (\bar{u} - u)$ done within the
statistical model, shows reasonably good agreement with the 
E615 data. Our values of $x v_{\pi}$ are, however, higher than those
due to the SMRS (dashed line) and GRS (dash-dotted line) 
parameterizations. Both these parameterizations have determined 
their valence densities by making a global fit to the $\pi N$
Drell-Yan data. To further compare our valence densities with
other calculations, we calculate the first two moments of the pion
valence quark distributions given by,
\begin{eqnarray}
<\,x\,v_{\pi}\,> \, = \, \int_0^1\, x\, v_{\pi}(x)\, dx \\
<\,x^2\,v_{\pi}\,> \, = \, \int_0^1\, x^2\, v_{\pi}(x)\, dx. 
\end{eqnarray}
In Fig. 1b we plot the $Q^2$ dependence of the pion valence moments. 
The solid curve corresponds to our calculation and the SMRS result
is shown by the hashed region. The points with error bars are the
results from a lattice QCD calculation \cite{latt} performed at
$Q^2=49$ GeV$^2$.

Next, we put the statistical model to test with the $\pi^- N 
\rightarrow \mu^+ \mu^- X$ Drell-Yan data. We calculate the
double-differential cross-section $d^2\sigma/dx_Fd\sqrt{\tau}$
where $x_F = x_{\pi} - x_N$ and $\sqrt{\tau} = x_{\pi} x_N =
M^2/s$. At leading order, $x_{\pi}$ and $x_N$ are the Bjorken $x$
variables of the pion and target nucleon respectively. M is
the invariant mass of the muon pair and $\sqrt{s}$ is the 
centre of mass energy. We calculate the Drell-Yan cross-sections
using the full next-to-leading order expressions \cite{smrs} and with
the choice $Q^2=M^2$. The proton PDFs required for the calculation
of the $\pi^- N$ Drell-Yan cross-sections are also calculated
within the statistical model. Both the pion and
nucleon PDFs are calculated using the same values of the FSC
parameters $a$ and $b$. A detailed description of the evaluation of
the nucleon PDFs can be found in ref.\cite{bhal}. To take into
account uncertainties due to normalization and higher order 
QCD corrections, we multiply our cross-sections by the standard
$K'$-factor. We analyse two different sets of data obtained from a 
$\pi^-$ beam incident on a tungsten target, $\pi^- W \rightarrow
\mu^+ \mu^- X$, by the NA10 and E615 collaboration.   
To make a correct comparison of our calculated cross-sections for
$\pi^- N \rightarrow \mu^+ \mu^- X$ with the $\pi^- W$ data, we
multiply our cross-sections with a factor $R = -0.55x_N +1.1$. This
form of the correction factor R is consistent with the observed
values \cite{rfac} of R as shown in ref.\cite{smrs}.
In Fig.2 we compare our calculated results with the NA10 data
taken at $\pi^-$ beam momenta of 194 and 286 GeV/c.
With $K'$ factors of 1.02 and 1.06 at beam momenta 194 and 286 GeV/c 
respectively, 
we find reasonably good agreement with
data at different values of $\sqrt{\tau}$. 
Our $K'$ values are close to those in 
ref.\cite{smrs} which range between 
1.1 and 1.4 depending on the type of fit.
A discussion on the $K'$ factor can be found in ref.\cite{alta}
where it was also noted that it should be fairly close to unity.
In Fig.2 we also show the dependence of the cross sections on 
the choice of the parameters $a$ and $b$. The solid curves correspond to
$a=-0.376$ and $b=0.504$ which are the values used throughout this work.
Varying the parameters arbitrarily we found that the agreement with
data reduces as we move away from the values of $a$ and $b$ quoted above.
To demonstrate this fact, we plot the cross sections at beam momentum 
286 GeV/c for two different sets of $a$ and $b$ in Fig. 2. The dashed
curves correspond to the calculations with $a=-0.2$, $b=0.3$ and the
dot-dashed curves correspond to those with $a=-0.6$, $b=0.7$.

In Fig.3 we show our calculations in comparison with the E615 data at
$P_L=252$ GeV/c. The $K'$ factor in this case is 0.85. It differs
from the $K'$ obtained by us for the NA10 data due to 
normalization errors in the data. Since the ratio $K'(NA10)/K'(E615)$
in this work is very similar to that obtained in 
refs.\cite{smrs,grs}, we do not worry about the fact that 
$K'(E615) < 1$. As mentioned above it could be due to errors in
normalization. The agreement of our 
results with data is again quite good. We are, however, not able to
reproduce data at very high or low values of $\sqrt{\tau}$ very well.
Although the reason for this disagreement is not clear to us, it certainly 
is not a short-coming of the statistical model. We say so because this
kind of disagreement was also found in ref.\cite{smrs}. They included
the data only in a certain range of $\sqrt{\tau}$ for fitting as they
were unable to reproduce the data at very high and low values of 
$\sqrt{\tau}$.

Finally in Fig. 4 we compare the valence, sea and gluon distributions
of the pion within the statistical model with those obtained
by the SMRS and GRS parameterizations. 
Since the pion sea distributions are not well determined by data, the
SMRS parameterization made different fits by varying the fraction of
the pion momentum carried by the sea between 10\% and 20\%. 
Interestingly, we find that the sea calculated
within the statistical model carries a fraction roughly equal to
15\% of the pion momentum which is like the SMRS best fit value.

To summarize, we can say that we have presented a simple 
physical model for the pion, using ideas from statistical mechanics.  
The only parameters in the
description of the pionic parton densities are the two `finite-size
correction' parameters. 
Considering the fact that the only two parameters appearing
in the model for the pion {\it have not been determined by any fit to
the pionic data}, the success of the model in reproducing data on
the pion is surprising. 
Since we use the same parameters for the
proton and pion (for reasons explained in the Introduction), 
we have in a sense, 
a unified model for the proton
and pion. We tried different sets of the values $a$ and $b$ for the 
pionic partons and having the same values as that for the proton is 
an outcome of our work rather than an assumption. 
At this stage, however, it would be too early to speculate about 
the significance of this
result. It could be that given a wealth of precise enough data on pionic 
parton distributions, the
`true' values of $a$ and $b$ determined by a fit, will differ 
from the ones presently used.
However, it is clear from our results that the difference will not be drastic.
It seems therefore that
the values for $a$ and $b$ are always close to the theoretical 
values, $a=-1/2$ and $b=3/2\pi$ determined in ref. \cite{paras}. 
This could partly be the reason why the statistical model works so well
for the proton as well as for the pion using the same $a$ and $b$.
Further efforts to understand why the ideas from 
statistical mechanics work inside hadrons could be worthwhile
and could improve our understanding of hadron structure. 
In future we intend to understand this point in more detail, by using for
comparison a larger set of data (i.e. prompt photon production),  
by extending the model
to other hadrons like kaons and hyperons, and relating 
their structure functions to fragmentation functions.  
This could also settle the issue about the
parameters $a$ and $b$.

\vskip0.5cm
\noindent
{\large \bf Acknowledgements} \\
One of the authors (NGK) gratefully acknowledges the warm hospitality
of the Physics Group at Universidad de Los Andes.

\newpage
\begin{figure}
\centerline{\vbox{
\psfig{file=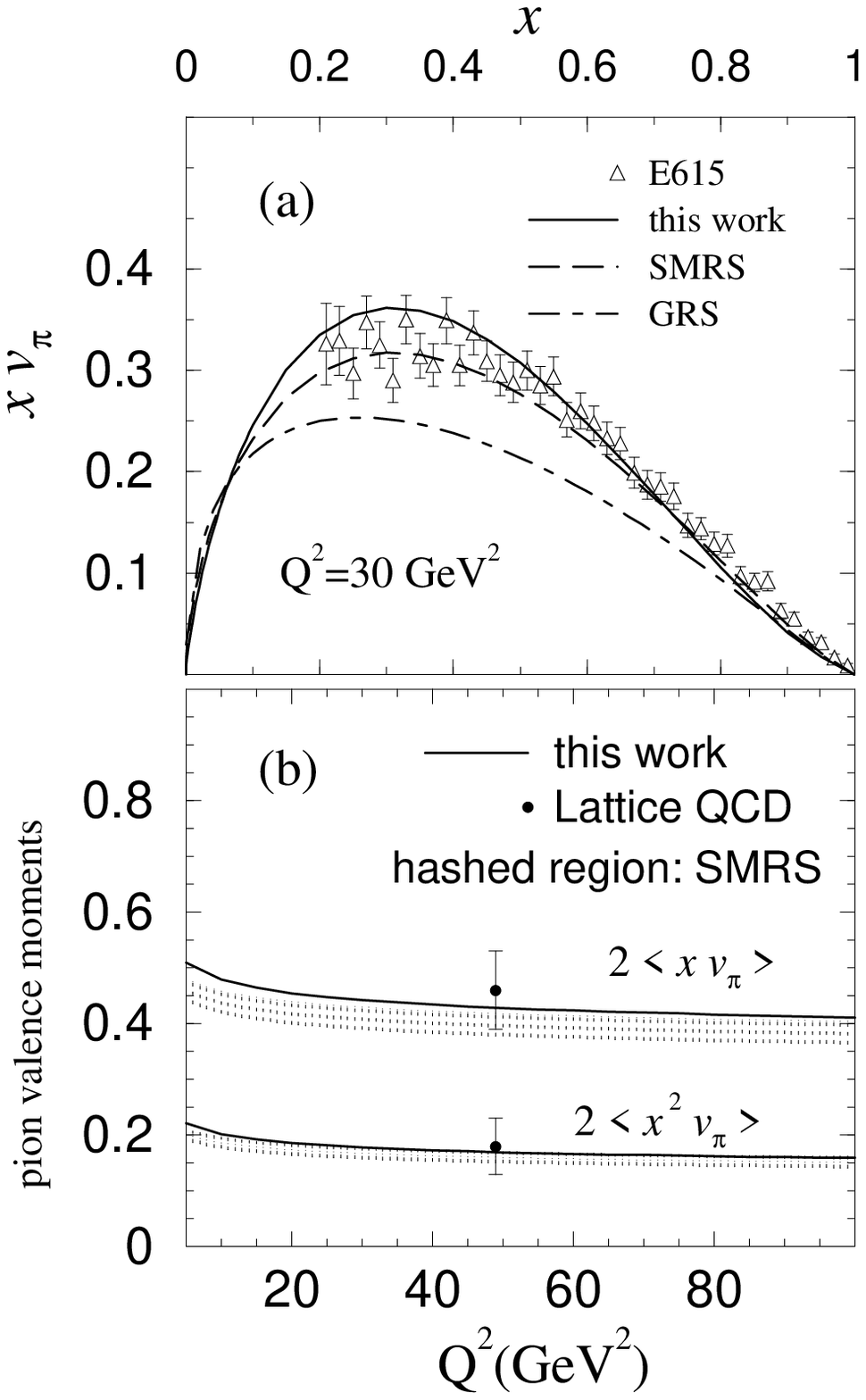,height=14cm,width=9cm}
}}
\caption{(a) Comparison of the valence structure function data from
ref.\cite{e615} with the valence distribution within the statistical
model (solid line). Dashed and dash-dotted lines are the valence
densities obtained by the SMRS (ref.\cite{smrs}) and GRS (ref.\cite{grs}) 
parameterizations respectively.
(b) The first two moments of the pion valence distribution as predicted
by the statistical model (solid line). Hashed region is the prediction 
\cite{smrs} from the fit to Drell-Yan data of NA10 and the filled circles
at $Q^2=49$ GeV$^2$ are lattice QCD \cite{latt} predictions.}   
\end{figure}
\newpage
\begin{figure}
\centerline{\vbox{
\psfig{file=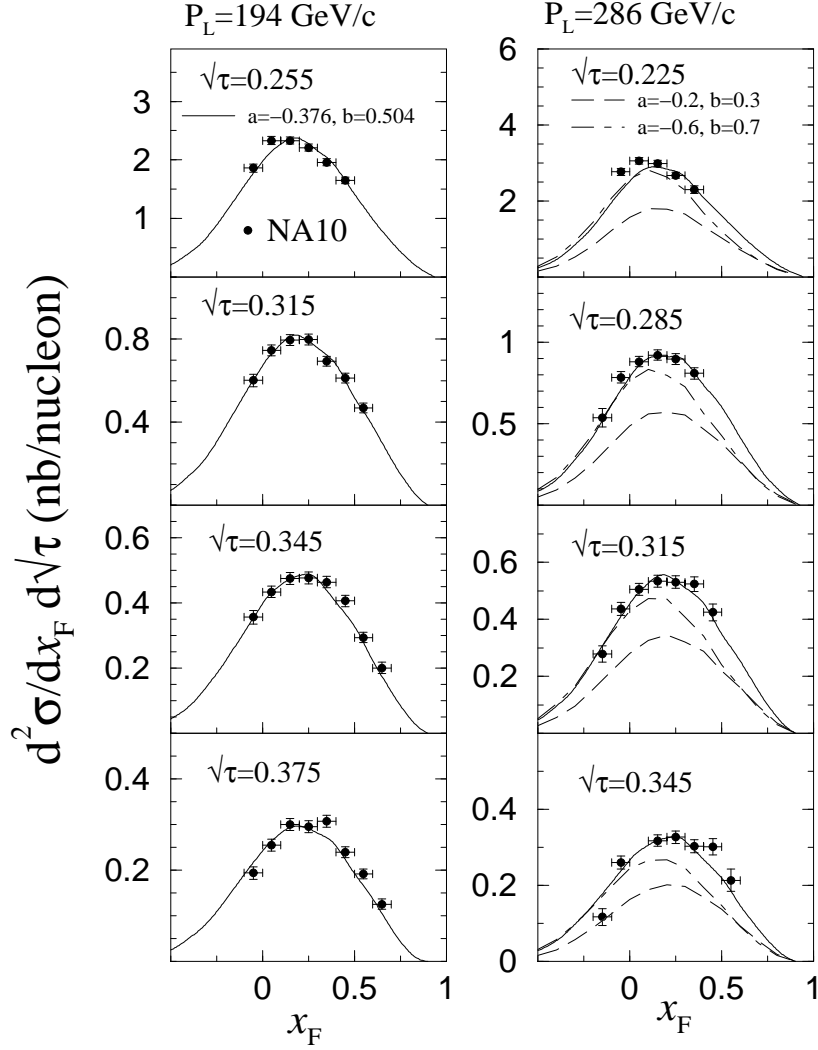,height=14cm,width=11cm}
}}
\caption{Drell-Yan data for $\pi^- W \rightarrow \mu^+ \mu^- X$
at beam momenta of 194 (left column) and 286 GeV/c (right column) taken by
the NA10 collaboration \cite{na10}. Solid lines are the  
next-to-leading order cross section predictions within
the statistical model with the finite-size correction parameters 
$a=-0.376$ and $b=0.504$ as used throughout this work. The dashed ($a=-0.2$, 
$b=0.3$) and 
dot-dashed lines ($a=-0.6$, $b=0.7$) at 286 GeV/c show the 
variation of the calculated cross 
sections with different choices of the parameters $a$ and $b$. 
The $K'$ factors for the 194 and 286 GeV/c data are
1.02 and 1.06 respectively.}
\end{figure}
\newpage
\begin{figure}
\centerline{\vbox{
\psfig{file=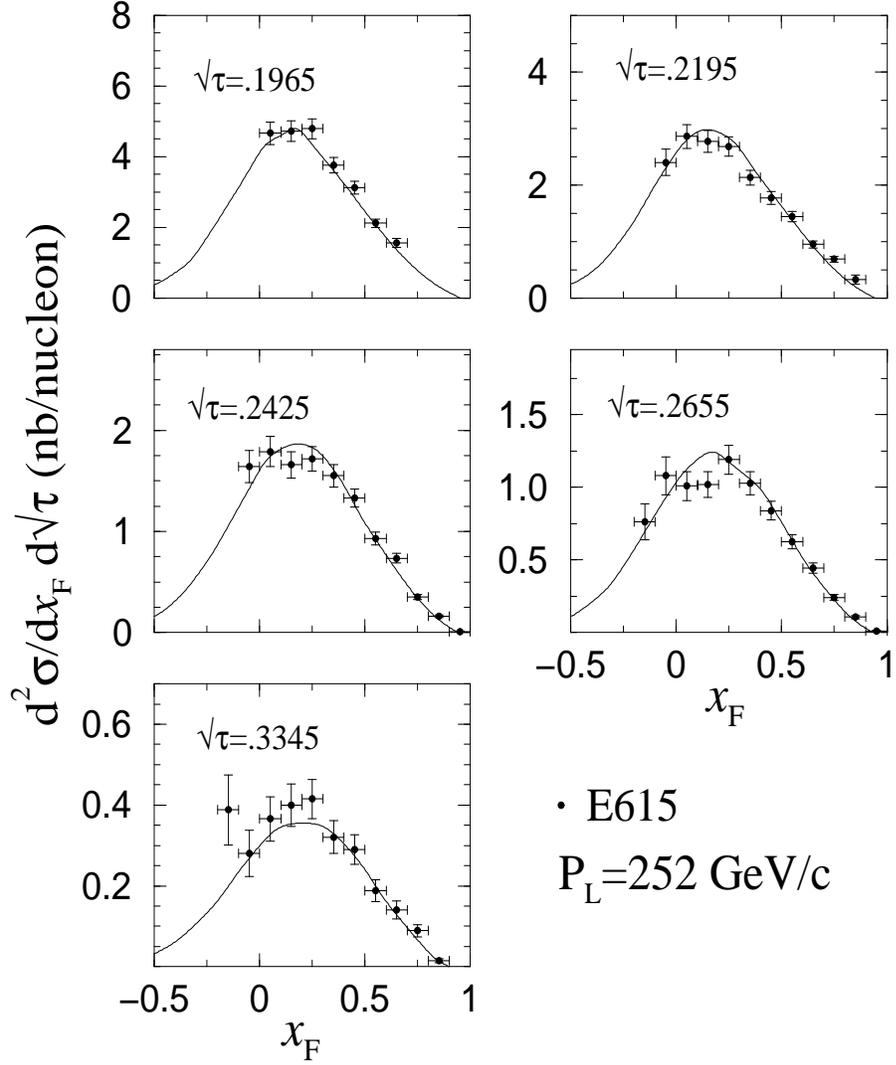,height=14cm,width=12cm}
}}
\caption{Drell-Yan cross section data for $\pi^- W \rightarrow \mu^+ \mu^- X$
at beam momentum 252 GeV/c taken by the E615 collaboration \cite{e615}. Solid
lines are the next-to-leading order cross section predictions within
the statistical model with a $K'$ factor of 0.85.}
\end{figure}
\newpage
\begin{figure}
\centerline{\vbox{
\psfig{file=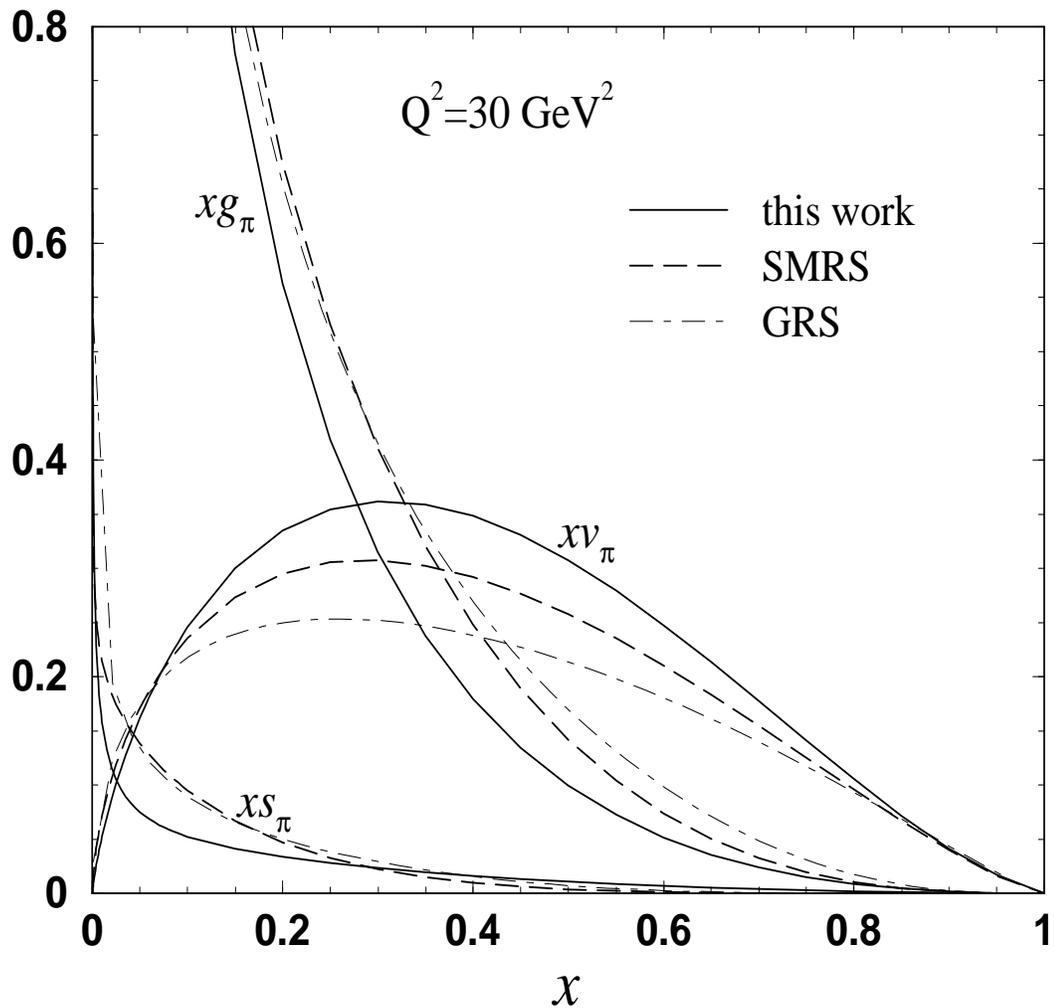,height=14cm,width=14cm}
}}
\caption{Comparison of the statistical model parton distributions 
(solid lines) with those due to the SMRS (ref.\cite{smrs}, dashed lines) and 
GRS (ref.\cite{grs}, dash-dotted lines) 
parameterizations at $Q^2 = 30$ GeV$^2$.}
\end{figure}
\end{document}